\documentclass[preprint]{elsarticle}
\usepackage{lineno}
\linenumbers

\usepackage{graphicx}
\usepackage{dcolumn}
\usepackage{bm}
\usepackage{hyperref}
\usepackage{amssymb}

\newcommand{\cma}{Cu$_2$MnAl }

\begin{document}

\title{Optical floating zone growth of high-quality Cu$_{2}$MnAl single crystals}

\author[1,2]{A. Neubauer\corref{cor1}}
\ead{andreas.neubauer@bam.de}

\author[1]{F. Jonietz}
\author[3]{M. Meven}
\author[3]{R. Georgii}
\author[3]{G. Brandl}
\author[4]{G. Behr}
\author[1]{P. B\"{o}ni}
\author[1]{C. Pfleiderer}
\cortext[cor1]{Corresponding author}
\address[1]{Physik Department E21, Technische Universit\"{a}t M\"{u}nchen, James Franck Stra\ss e, 85748 Garching, Germany}
\address[2]{Bundesanstalt f\"ur Materialforschung und -pr\"ufung, 12205 Berlin, Germany}
\address[3]{Forschungsneutronenquelle Heinz Maier Leibnitz (FRM II),  Lichtenbergstr. 1, 85748 Garching, Germany}
\address[4]{IFW Dresden, PF 270116, D-01171, Dresden, Germany}
\date{\today}

\begin{abstract}
We report the growth of large single-crystals of Cu$_{2}$MnAl, a ferromagnetic Heusler compound suitable for polarizing neutron monochromators, by means of optical floating zone under ultra-high vacuum compatible conditions. Unlike Bridgman or Czochralsky grown Cu$_{2}$MnAl, our floating zone grown single-crystals show highly reproducible magnetic properties and an excellent crystal quality with a narrow and homogeneous mosaic spread as examined by neutron diffraction.  An investigation of the polarizing properties in neutron scattering suggests a high polarization efficiency, limited by the relatively small sample dimensions studied. Our study identifies optical floating zone under ultra-high vacuum compatible conditions as a highly reproducible method to grow high-quality single-crystals of Cu$_{2}$MnAl.
\end{abstract}

\begin{keyword}
single crystal growth \sep optical floating zone \sep polarized neutron scattering \sep polarizing monochromator \sep mosaicity 
\end{keyword}

\maketitle

\section{\label{sec:level1}Introduction}

Heusler compounds exhibit a remarkably wide variety of different electronic ground states ranging from simple metallic, over semiconducting to insulating behavior including recent theoretical proposals, which suggest the possibility of topological insulators\,\cite{cha10}. Heusler compounds also stabilize various forms of electronic order including half-metallic ferromagnetism and superconductivity\,\cite{graf2011}. This has motivated great efforts to exploit the wide range of ground states by combining different materials in 'all-Heusler devices'\,\cite{fel09, spr10, erb10}. Based on these activities the preparation of high-quality single crystals of Heusler compounds  is of considerable general interest.

In this paper we address the properties of Cu$_2$MnAl, the first compound made of non-ferromagnetic elements, in which Fritz Heusler discovered ferromagnetism in 1903\,\cite{heu03}. {\cma} orders ferromagnetically at $T_{\rm C}$\,=\,622\,K with an ordered magnetic moment of 3.6\,$\mu_{\rm B}$ per formula unit\,\cite{mic78}. {\cma} is mostly known for its use in polarizing monochromators for neutron scattering techniques\cite{wil88}. 

Scattering off the $(111)$ Bragg peak is thereby typically used to generate a monochromatic polarized neutron beam\,\cite{del71, fre83}. In turn, the main challenge in the preparation of monochromators for polarized neutron scattering using Cu$_{2}$MnAl Heusler single crystals consists in the growth of large and homogeneous crystals with a well defined mosaic spread. A mosaic spread in the range of $0.2^{\circ}-1^{\circ}$ is desirable in order to match the divergences of the neutron beam, and hence obtain large intensities \cite{fre09}. As the main problem previous studies established that {\cma} single-crystals prepared by the Bridgman technique are characterized by very large, uncontrolled anisotropies of the mosaic distribution depending on the growth direction\cite{cou99, cou04}. Due to this sensitivity of the crystal quality on the growth conditions less than 50\% of the crystals are suitable for monochromators. Moreover, it implies the need for very careful screening of the samples to identify sections of the ingots with suitable properties. To highlight the need for advances in the synthesis of Heusler single crystals one has to consider that modern focusing and double-focusing monochromator devices often need in the order of 100 crystals or more.  

In this paper we report the growth of single crystals of Cu$_{2}$MnAl by means of optical floating zone. Altogether eight single crystals were grown, two in a vertical double ellipsoid image furnace at IFW Dresden and six in a UHV-compatible four-mirror image furnace at TU Munich\,\cite{neu11}. The UHV compatible conditions were found to promote stable growth conditions, resulting in a mono-crystalline state over the entire cross-section of the rods. The magnetic properties and the crystal structure of four of those six crystals were investigated in detail. In addition, a systematic study of the polarizing properties of these crystals in neutron scattering was carried out at the diffractometer MIRA at FRM\,II. To assess the properties of the floating zone grown crystals they were compared with commercial Bridgman grown Cu$_{2}$MnAl crystals investigated in the same way. As our main result we find that optical floating zone growth under UHV compatible conditions is ideally suited to grow high quality single crystals of {\cma} with a small isotropic mosaic spread.

\section{\label{sec:level2}Single crystal growth}

Single-crystal growth of {\cma} is challenging, as the detailed ternary phase diagram of the Cu-Mn-Al system has not been reported. Yet, a series of publications suggests that {\cma} is congruently melting and crystallizes around 1125\,K in the cubic $\rm{L2_1}$ Heusler structure\,\cite{mic78, cou99, kos66, dub79, sak90, kai98}. This so-called $\beta$\,-\,phase appears to be metastable and does not exist in the equilibrium phase diagram at room temperature. Below around 923\,K {\cma} presumably decomposes in a solid state reaction into Cu$_9$Al$_4$, Cu$_3$Mn$_2$Al, and $\beta$\,-\,Mn phases. However, since the transformation kinetics of this solid state reaction slows down dramatically well below 923\,K, it has been possible to prepare {\cma} in a quasi-stable state at room temperature when cooling samples sufficiently fast. The hidden agenda in our study was hence, whether the temperature gradient along the sample in optical-float zoning may drive such a decomposition.

\begin{figure}[t]
\includegraphics[width=\columnwidth]{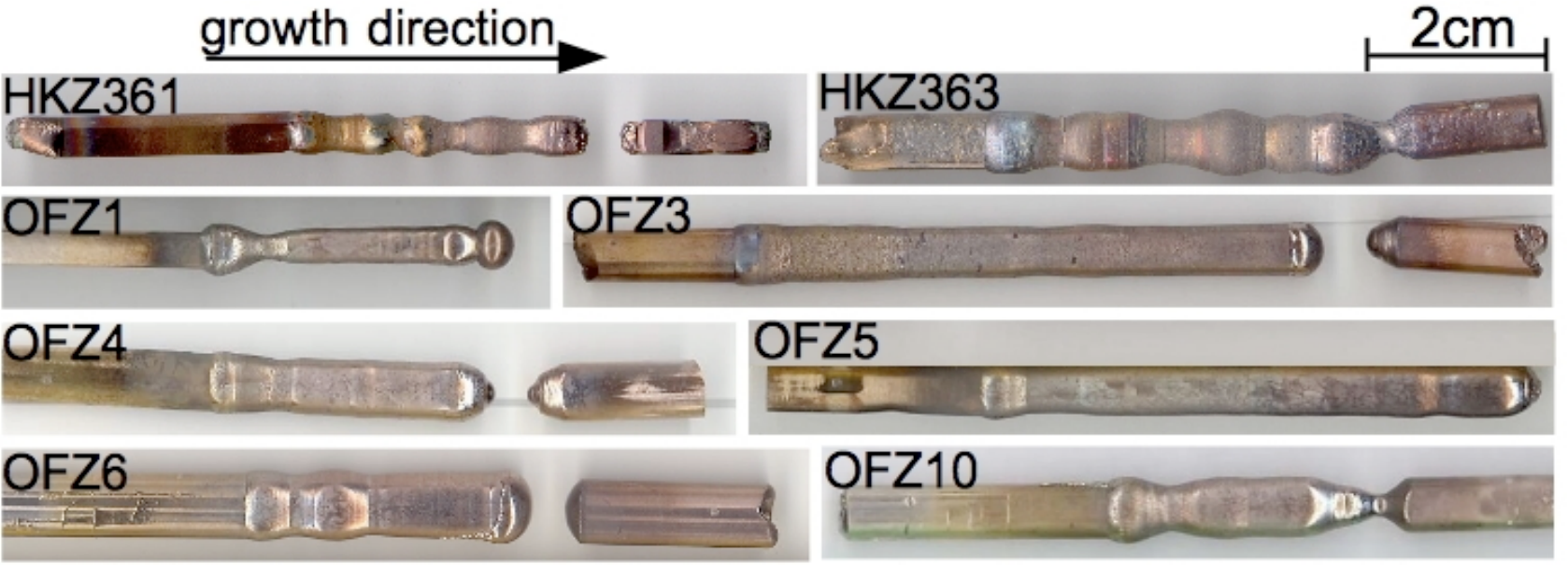}
\caption{The Cu$_{2}$MnAl crystals grown by optical floating zone for this study. HKZ361 and HKZ363 were grown at the IFW in Dresden. An abundance of oxide contamination disturbed stable growth conditions, leading to repeated separations of the zone during crystal growth. The inhomogeneous shape of the crystals (swellings and contractions) and the gray and brown staining are the visible results. In comparison, the Cu$_{2}$MnAl crystals OFZ1, OFZ3, OFZ4, OFZ5, OFZ6, and OFZ10 were grown with a UHV-compatible image furnace at the TUM\cite{neu11}, leading to stable growth conditions. A more homogeneous shape of the rods and a reduced staining are the visible results.  The length scale shown in the top right applies to all crystals.}
 \label{ TUMIFW}
\end{figure}

Shown in Fig.\,\ref{ TUMIFW} are the eight crystals grown for our study by vertical floating zone. Two crystals were grown in a vertical double ellipsoid image furnace (model URN-2-ZM, MPEI, Moscow) at the IFW in Dresden. They are labeled HKZ361 and HKZ363. The other six crystals were grown in a refurbished UHV-compatible four-mirror image furnace (model CSI FZ-T-10000-H-III-VPS) at the Technical University in Munich\,\cite{neu11}. They are labeled OFZ1, OFZ3, OFZ4, OFZ5, OFZ6, and OFZ10. 

Rectangular bars with a rectangular cross-section of $4\,\times\,4\,{\rm mm^{2}}$ prepared from stoichiometric Bridgman grown single crystals were used as starting rods for the floating zone growth of HKZ361 and OFZ1. For all other crystals cylindric seed and feed rods of stoichiometric Cu$_{2}$MnAl polycrystals with a diameter of 6\,mm were prepared in bespoke rod-casting furnace at the IFW Dresden and TUM, respectively \cite{neu12}.  The rod-casting furnace at TUM was especially designed to offer UHV-compatible conditions \cite{neuphd11,mun09,bau09}.

Crystals HKZ361 and HKZ363 were grown at a rate of 15\,mm/h and with a counter-rotation of 40\,rpm (seed) and 25\,rpm (feed). Growth took place in a high purity flowing Argon gas environment (6\,-\,10\,l/h) at $p\sim1.1$\,bar. An abundance of oxide contamination flowing on the molten zone disturbed a stable growth process for both crystals. Attrition of oxide layers led to a shaking of the zone that resulted in repeated separations of the zone during the growth process. The resulting inhomogeneous shape of the crystals and the strong contamination with oxide on the outside of HKZ361 and HKZ363 (indicated by the grey and brown staining) can be seen in  Fig.\,\ref{ TUMIFW}. 

\begin{figure}[t]
\centering
\includegraphics[width=0.4\columnwidth]{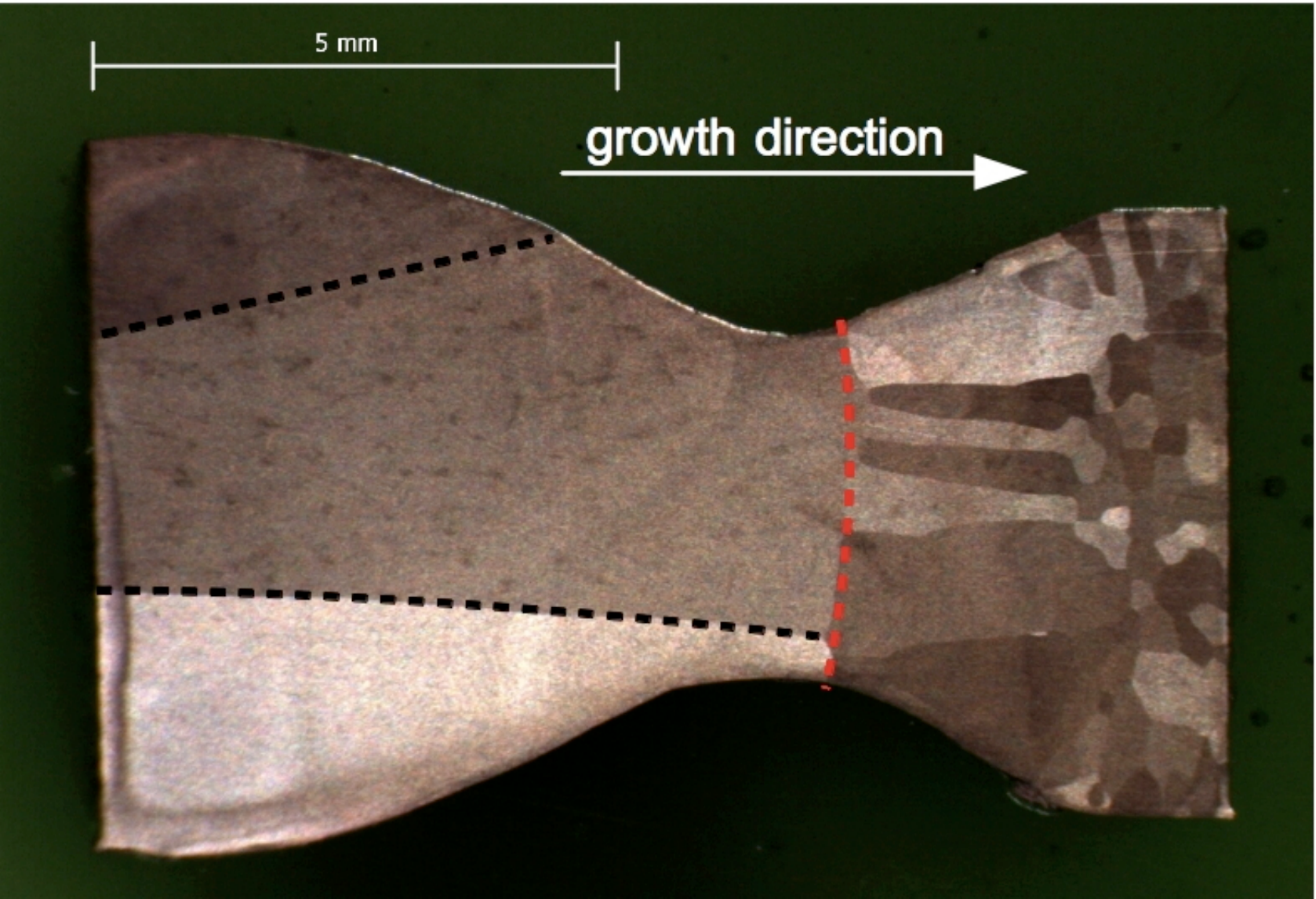}
 \caption{(a) Final zone of HKZ363. Three large grains can be clearly identified on the left hand side of the interface (indicated by the horizontal dashed lines). These grains are separated by a slightly convex growth interface (see vertical dashed line) from the poly-crystalline feed rod, where large grains already form probably due to annealing.  The surfaces was polished and etched with Marble reagent (2.5\,g CuSO$_{4}$\,+\,30\,cm$^{3}$\,HCl\,+\,25\,cm$^{3}$\,H$_{2}$O) in order to highlight grain structures.}
 \label{fig:mik}
\end{figure}

Large single crystal grains formed in all crystals as illustrated in Fig.\,\ref{fig:mik}. The image shows the quenched last zone of HKZ363. The surface was etched with Marble reagent (2.5\,g CuSO$_{4}$\,+\,30\,cm$^{3}$\,HCl\,+\,25\,cm$^{3}$\,H$_{2}$O) for a better visibility of the grain structure. The growth direction is from the left to the right. In the crystal three grains can be identified with the grain at the center expanding in size. A slightly convex growth interface (marked by the vertical dashed line) separates the crystal that was grown from the poly-crystalline structure of the feed rod. In the poly-crystalline feed large grains already formed in the vicinity of the growth interface due to annealing. EDX investigation of the surface showed a stoichiometric Cu$_{2}$MnAl composition with no indication of secondary phases.  These findings are consistent with earlier reports of Cu$_{2}$MnAl as a congruently melting compound that shows a strong trend to crystallize in a mono-crystalline state.

Crystals OFZ1, OFZ3, OFZ4, OFZ5, and OFZ6 were grown in the UHV-compatible image furnace at TUM at growth rates in the range 10-12\,mm/h. In contrast, for OFZ10 the growth rate was increased from 5\,mm/h to 10\,mm/h during the growth (we return to this issue later). In each growth process the feed and seed rod were counter-rotating with 10\,rpm and 30\,rpm, respectively. Prior to each growth process the image furnace was carefully baked ($10^{-8}\,{\rm mbar}$) and filled with 6N Argon gas, that was additionally purified with a getter furnace. Each growth process took place in a static Argon atmosphere of $p\,\sim\,1.5\,{\rm bar}$. A strong reduction of the oxide layers floating on the molten zone was observed in comparison to the crystals grown in the non-UHV compatible furnace at IFW Dresden. This is clearly illustrated by the difference in surface contamination shown in Fig.\,\ref{ TUMIFW}. For the high-purity environment a stable molten zone formed readily during the whole growth process. We attribute the complete grain selection process, that resulted in a mono-crystalline structure across the entire cross section of the rod for all crystals grown in UHV-compatible image furnace at TUM to this improved stability of the zone.

\begin{figure}[t]
\centering
\includegraphics[width=0.4\columnwidth]{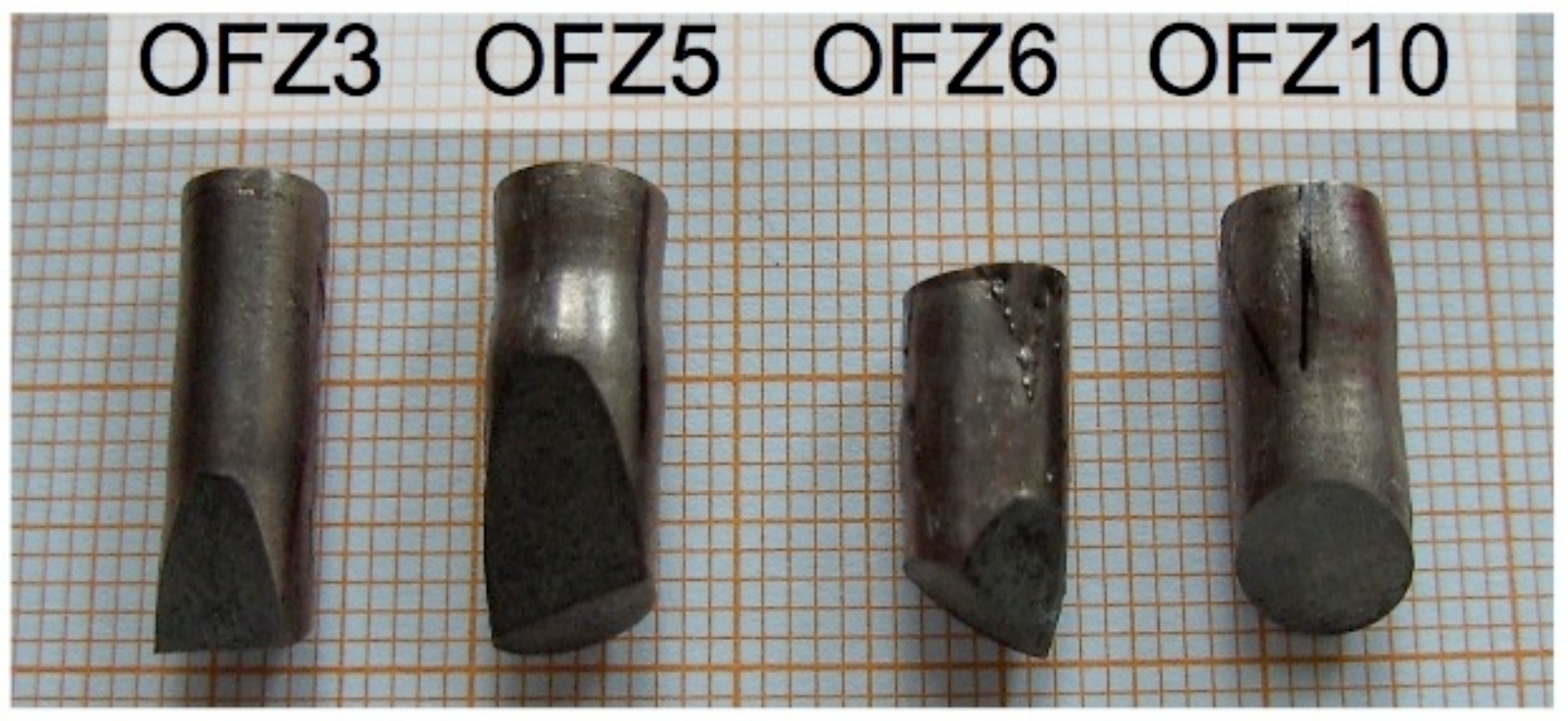}
 \caption{(a) Single-crystalline Cu$_{2}$MnAl crystals investigated in detail. No preferred growth direction of the crystal structure could be identified.}
 \label{fig-Heidi-Messungen}
\end{figure}

For studies of the magnetization and neutron scattering large single-crystalline samples were prepared from OFZ3, OFZ5, OFZ6 and OFZ10, as shown in  Fig.\,\ref{fig-Heidi-Messungen}. Single-crystallinity of the samples was at first established with a light microscope and by means of x-ray Laue diffraction. The orientation of the crystal structure with respect to the growth direction was determined by means of x-ray Laue diffraction.  A different crystal orientation was found for each crystal. Since poly-crystalline rods rather than oriented crystals were used as seed rods in each growth process, this means that no preferred growth direction could be identified. In turn, this suggests that oriented seed crystals may allow to grow cylindrical single crystals of arbitrary crystallographic orientation. 

\section{\label{sec:level3}Magnetization}

The ferromagnetic properties of the Cu$_{2}$MnAl single-crystals grown served as first test of the sample quality. In order to avoid systematic errors due to demagnetizing effects in Cu$_{2}$MnAl\,\cite{fre83} and to be able to quantitatively compare the magnetic properties, oriented samples of the same dimensions were prepared from OFZ3, OFZ5, OFZ6, OFZ10 and from a Bridgman grown single crystal (BM). The samples were cut in a rectangular parallelepiped 5\,$\times$\,2.5\,$\times$\,2\,mm$^{3}$ (cf. Fig.\,\ref{fig:ProbenMira}) with the flat front surface being a (111) crystallographic plane and the small bottom surface a (110) plane. 
\begin{figure}[h]
\centering
\includegraphics[width=0.5\columnwidth]{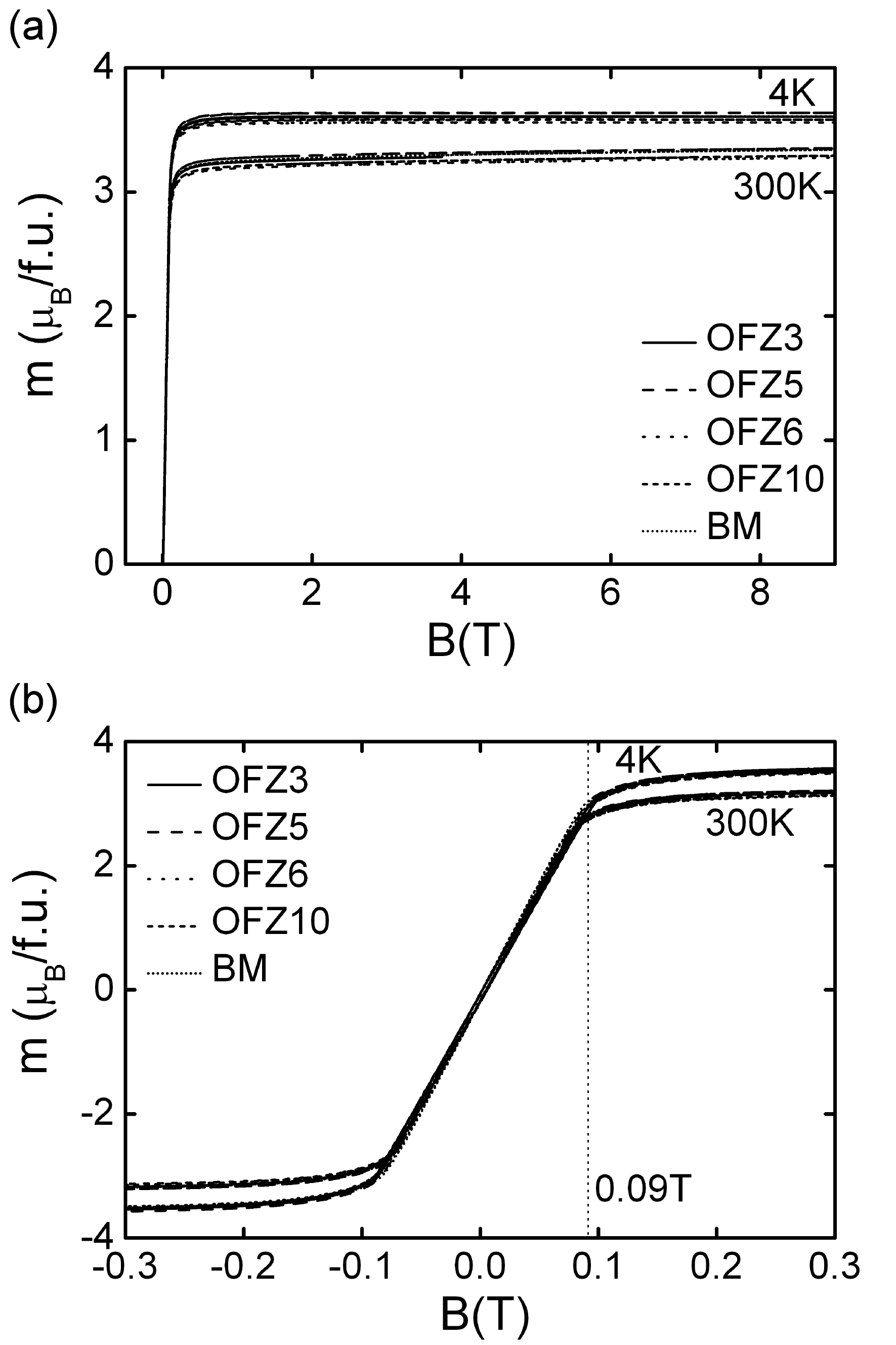}
 \caption{(a) Magnetization as a function of magnetic fields up to 9\,T at $T$\,=\,4\,K and $T$\,=\,300\,K. All floating zone grown Cu$_{2}$MnAl samples, OFZ3, OFZ5, OFZ6, and OFZ10 as well as the Bridgman grown sample (BM) show the same behavior and saturate at $m$\,$\sim$\,3.6\,$\mu_{\rm{B}}$/f.u. and $m$\,$\sim$\,3.2\,$\mu_{\rm{B}}$/f.u. at $T$\,=\,4\,K and $T$\,=\,300\,K, respectively. (b) Magnetization as a function of magnetic fields for low fields. All samples saturate ferromagnetically above $B$\,$\geq$\,90\,mT. Data are not corrected for demagnetizing effects.}
 \label{fig:Cu2MnAl-MVergleich}
\end{figure}

All five samples were measured in a vibrating sample magnetometer (VSM) with the magnetic field parallel to the long axis of the parallelepiped, i.e., the configuration that minimizes the demagnetizing effects. Figs.\,\ref{fig:Cu2MnAl-MVergleich} (a) and (b) show the field dependent magnetization at 4\,K and 300\,K for fields up to 9\,T and for low fields, respectively. At high fields and 4\,K the magnetization saturates for all samples at $m\,\sim\,3.6\,\mu_{\rm{B}}$/f.u.. At 300\,K the magnetization saturates for all samples at  $m\,\sim\,3.2\,\mu_{\rm{B}}$/f.u.. Both values are in excellent agreement with the literature\,\cite{mic78}. At low fields the magnetization shows a linear slope followed by the onset of saturation at 90\,mT for both temperatures and all samples. No evidence suggesting hysteretic behavior is observed in any of the samples. The  magnetic moments of the samples differ by less than 3\,\%. Hence, the magnetic properties of the floating zone grown crystals are in excellent agreement with respect to each other and as compared with the Bridgman grown sample. This establishes highly reproducible ferromagnetic properties of the floating zone grown crystals. 

\section{\label{sec:level4}Single crystal neutron diffraction}

In order to investigate the mosaic spread of the floating zone grown crystals, neutron scattering experiments at the single crystal diffractometer HEIDI\,\cite{heidi} at FRM\,II were carried out. Neutrons with a wavelength of $\lambda=0.87$\,{\AA} (Cu-220 monochromator) were used with a primary collimation of 30$^{\prime}$. Single crystals of different dimensions  prepared from OFZ3, OFZ5, OFZ6 and OFZ10 (see Fig.\,\ref{fig-Heidi-Messungen}) were investigated as well as a Bridgman grown (BM) single-crystalline plate with dimensions 20$\times$40$\times$3\,mm$^{3}$.

For each crystal rocking scans with respect to the $\{400\}$ and $\{111\}$ lattice planes were carried out. For OFZ10 the $\{333\}$ lattice planes were studied and for BM a single (333) plane. Both the integrated and absolute intensities of the Bragg reflections of the different rocking scans vary because different sample volumes were measured for each direction. Nevertheless, the crystal mosaicity was obtained from the width of the rocking curves in terms of the full-width-have-maximum (FWHM) and taking into account the resolution function of the instrument\,\cite{res}. 

\begin{figure}[p]
\centering
\includegraphics[width=\columnwidth]{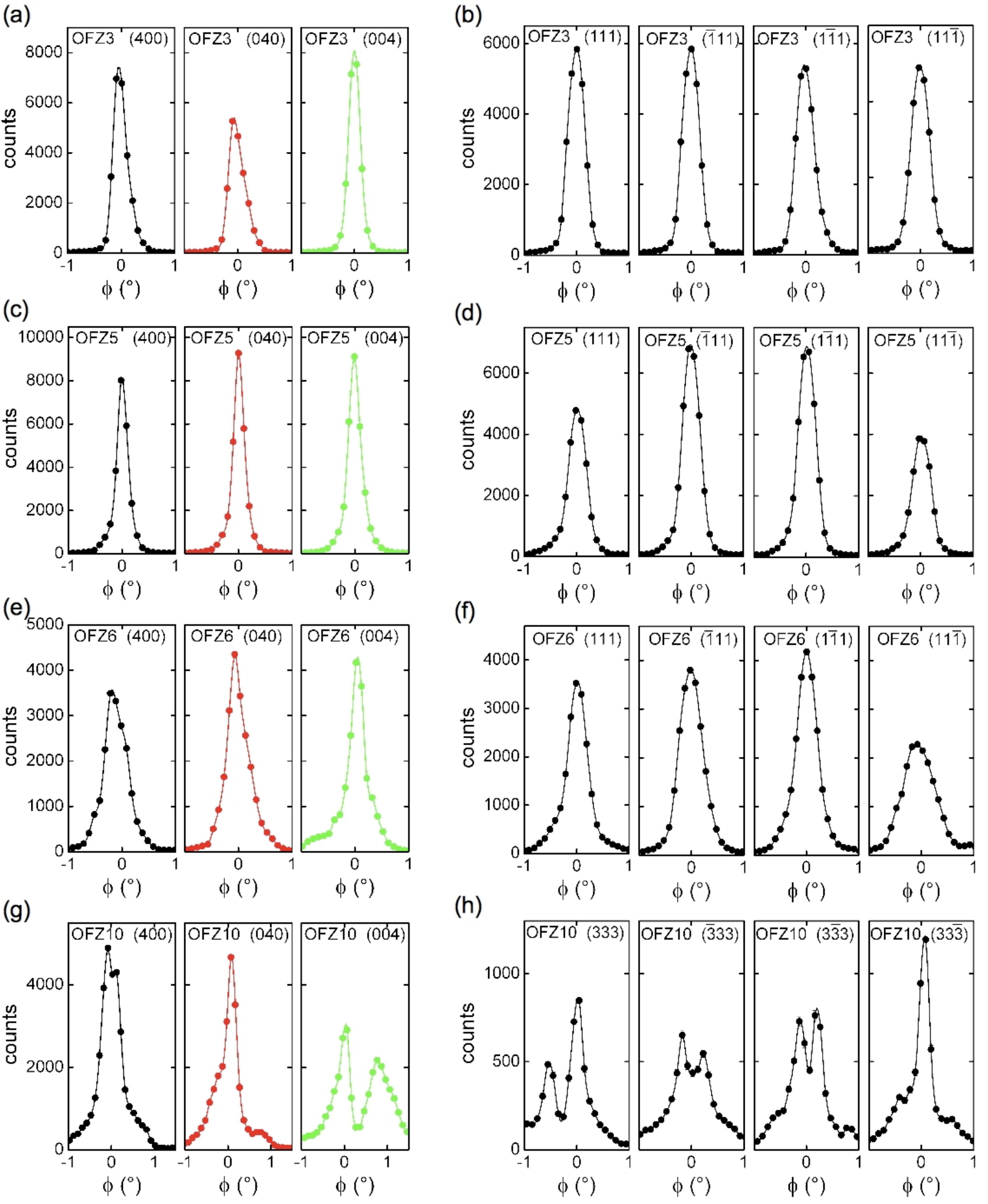}
 \caption{Overview of the $\{400\}$ and $\{111\}$ Bragg scattering intensities as a function of the rocking angle $\phi$ for the floating zone grown crystals. OFZ3 (a,\,b) and OFZ5 (c,\,d) show highly homogeneous shaped rocking curves for all $\{400\}$ and $\{111\}$ planes. For OFZ6 (e,\,f) the curves are slightly broadened. The inhomogeneous peak structure of crystal OFZ10 (g,\,h) is most likely due to a change of the growth rate during crystal growth. However, one has to take into consideration that the instrumental resolution for $\{333\}$ is better than for $\{111\}$ \cite{res}, leading to the more narrow peak structure in (h). The step width of the rocking scans was $0.1^{\circ}$.}
 \label{fig:Heidi-overview}
\end{figure}

An overview of the $\{400\}$ and $\{111\}$ Bragg scattering intensities as a function of the rocking angle $\phi$ is shown in  Fig.\,\ref{fig:Heidi-overview}. OFZ3 (a,\,b) and OFZ5 (c,\,d) show similar, highly homogeneously shaped rocking curves for all $\{400\}$ and $\{111\}$ reflections. This is confirmed by the very homogeneous mosaic distribution around 0.25$^{\circ}$ for OFZ3 and OFZ5, respectively.  In comparison, the accuracy of measurement was $\pm 0.05^{\circ}$. The mosaicities are summarized in Table\,\ref{tab:heidi}. The rocking curves for OFZ6 (e,\,f) are slightly broadened due to a small second peak. This deviation also shows up in terms of the larger anisotropy of the mosaicity for the different scattering planes. Nevertheless, in comparison to the data reported for the Bridgman grown crystals, OFZ6 shows an essentially isotropic mosaic distribution.

Clear deviations from an isotropic mosaic spread are found for OFZ10 (g,\,h), where two intensity maxima are seen for most of the reflections. This signature is most likely due to the use of two different growth velocities (10\,mm/h and 5\,mm/h) during the floating zone growth of OFZ10. This sensitivity of the mosaic distribution to variations of the growth rate might be advantageous when growing crystals with a given mosaic spread for use as polarizing neutron monochromators. As mentioned above, a mosaic spread (0.2$^{\circ}$\,$-$\,1.0$^{\circ}$) is necessary for high neutron intensities. 

\begin{figure}[t]
\centering
\includegraphics[width=0.9\columnwidth]{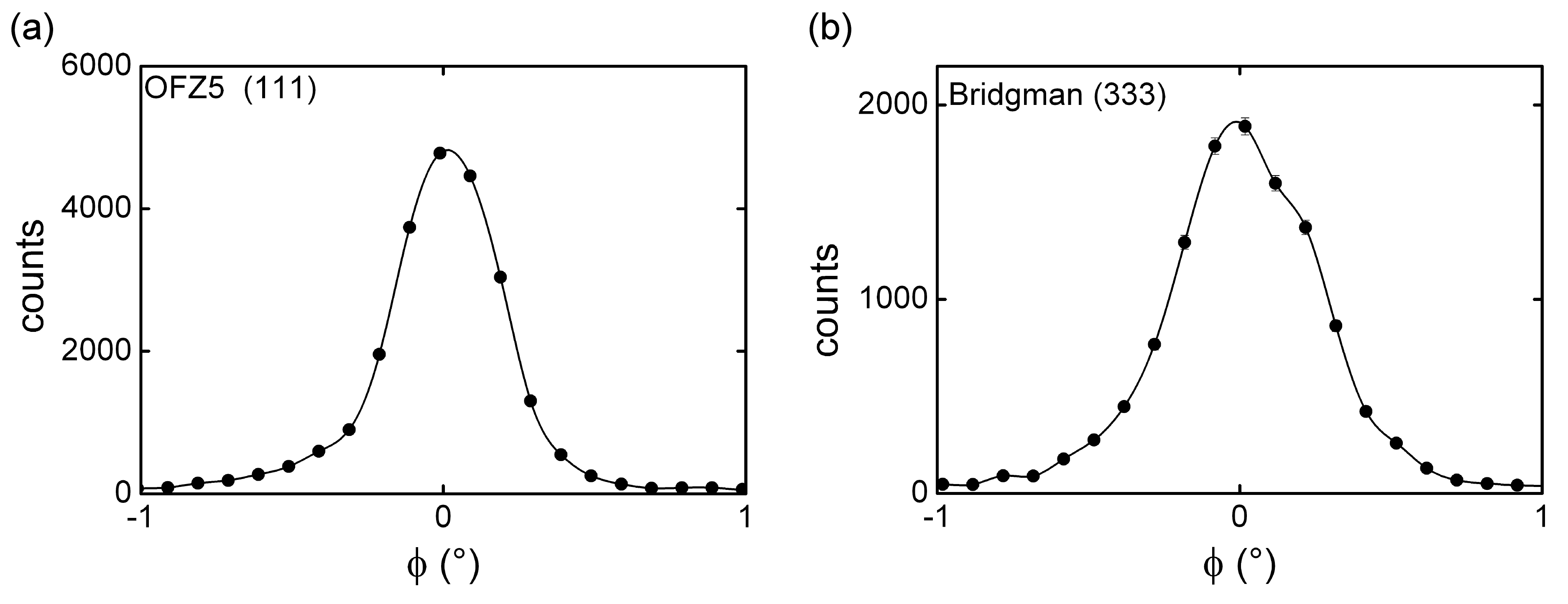}
 \caption{Comparison of the Bragg scattering intensities of OFZ5 (111) and the large Bridgman grown plate (333) as a function of the rocking angle $\phi$. The rocking curve of the Bridgman crystal is slightly broadened, indicating a coarser mosaic spread in comparison with the floating zone grown crystal OFZ5.}
 \label{fig:Bridgman}
\end{figure}

Fig.\,\ref{fig:Bridgman} shows the rocking scan of the (111) plane of OFZ3 and the (333) plane of the large Brigdman grown sample.  In comparison to the floating zone grown crystal,  the rocking scan of the Bridgman grown crystal is less homogeneous and has a slightly larger mosaicity. The comparison shows that optical floating zone allows to reproducibly grow Cu$_{2}$MnAl single crystals with a \textit{homogeneous} mosaic spread that is at least comparable to the mosaic spread of  ``good'' Bridgman grown crystals. 

\begin{table*}[h]
\begin{center}
\begin{tabular}{|p{3.5cm}|p{1.1cm}|p{1.1cm}|p{1.1cm}|p{1.1cm}|p{1.1cm}|}
\hline
Mosaicity \textbackslash  Crystal&OFZ3&OFZ5&OFZ6&OFZ10&BM\\
\hline\hline
(400)\newline (040)\newline (004) & 0.25$^{\circ}$\newline 0.25$^{\circ}$\newline 0.22$^{\circ}$ & 0.21$^{\circ}$\newline0.22$^{\circ}$\newline0.25$^{\circ}$ & 0.48$^{\circ}$\newline0.40$^{\circ}$\newline0.29$^{\circ}$ & 0.50$^{\circ}$\newline0.30$^{\circ}$\newline1.27$^{\circ}$&\\
\hline
 (111) \newline ($\overline 111$) \newline ($1\overline11$) \newline ($11\overline1$)& 0.24$^{\circ}$\newline 0.20$^{\circ}$\newline 0.25$^{\circ}$\newline 0.25$^{\circ}$ & 0.23$^{\circ}$\newline0.22$^{\circ}$\newline0.23$^{\circ}$\newline 0.27$^{\circ}$ & 0.27$^{\circ}$\newline0.41$^{\circ}$\newline0.31$^{\circ}$ \newline 0.64$^{\circ}$ & &\\
\hline
 (333) \newline ($\overline 333$)\newline ($3\overline33$) \newline ($33\overline3$)& 0.18$^{\circ}$\newline 0.21$^{\circ}$\newline 0.26$^{\circ}$\newline 0.27$^{\circ}$ &  &  & 0.73$^{\circ}$\newline0.71$^{\circ}$\newline0.64$^{\circ}$\newline 0.23$^{\circ}$& 0.53$^{\circ}$\\
\hline
\end{tabular}
\end{center}
\caption{Crystal mosaicities for different scattering planes of the floating zone (OFZ3, OFZ5, OFZ6 and OFZ10) and Bridgman grown (BM) Cu$_{2}$MnAl crystals. The mosaicities were calculated from the FWHM values of the Bragg peaks taking into account the instrumental resolution function\,\cite{res}. The accuracy of the mosaicities is better than $\pm$\,0.05$^{\circ}$.}
\label{tab:heidi}
\end{table*}

\section{\label{sec:level5}Polarizing properties}

We finally turn to the polarizing properties of the floating zone grown single crystals.  As reported in the literature\,\cite{del71, fre83}, the Bragg (111) reflection of Cu$_{2}$MnAl may be used to generate a monochromatic beam of polarized neutrons. In general, scattering of an unpolarized neutron beam on a ferromagnet results in individual structure factors for nuclear $F_{\rm nuc}$ and magnetic $F_{\rm mag}$ scattering, that sum up individually to a common scattering intensity\,\cite{bac53,web69}. For a ferromagnet the scattering intensity is given as
\begin{equation}
I\propto F^{2}_{\rm{tot}}= F^{2}_{\rm{nuc}}+q^{2}F^{2}_{\rm{mag}},
\label{Izf}
\end{equation}
where $\textbf{q}$ is the magnetic interaction vector and $q^{2}$\,=\,$\sin^{2}\alpha$. $\alpha$ is the angle between the magnetization and the scattering vectors. For a cubic magnetic crystal without anisotropy, as it is the case for Cu$_{2}$MnAl, and in an unsaturated magnetic state $q^{2}$ takes a value of 2/3. 

For a saturated ferromagnet with the magnetization direction perpendicular to the scattering vector, $q^{2}$ takes a value of 1. In this case the scattering intensity for neutrons with spin parallel ($+$) to the magnetization direction is given as 
\begin{equation}
I_{\rm{+}}\propto F^{2}_{\rm{tot,+}}=F^{2}_{\rm{nuc}}+F^{2}_{\rm{mag}}, 
\label{Iplus}
\end{equation}
while
\begin{equation}
I_{\rm{-}}\propto F^{2}_{\rm{tot,-}}=F^{2}_{\rm{nuc}}-F^{2}_{\rm{mag}}
\label{Iminus}
\end{equation}
is the scattering intensity for neutrons with spin antiparallel ($-$) to the magnetization direction. 

In the case of Cu$_{2}$MnAl the magnetic structure factor for (111) Bragg scattering is comparable to the nuclear structure factor, i.e., $F_{\rm{mag}}^{(111)} \simeq F_{\rm{nuc}}^{(111)}$\,\cite{del71}. Considering Eq.\,\ref{Iplus} and Eq.\,\ref{Iminus}, scattering of the (111) plane hence leads to a high flipping ratio $R=I_{+}/I_{-}$ and a polarized neutron beam with a high polarization $P$ defined as
\begin{equation}
P = \frac{I_{+}-I_{-}}{I_{+}+I_{-}}.
\end{equation}

\subsubsection{Experimental set-up}
\begin{figure}[b!]
\centering
\includegraphics[width=\columnwidth]{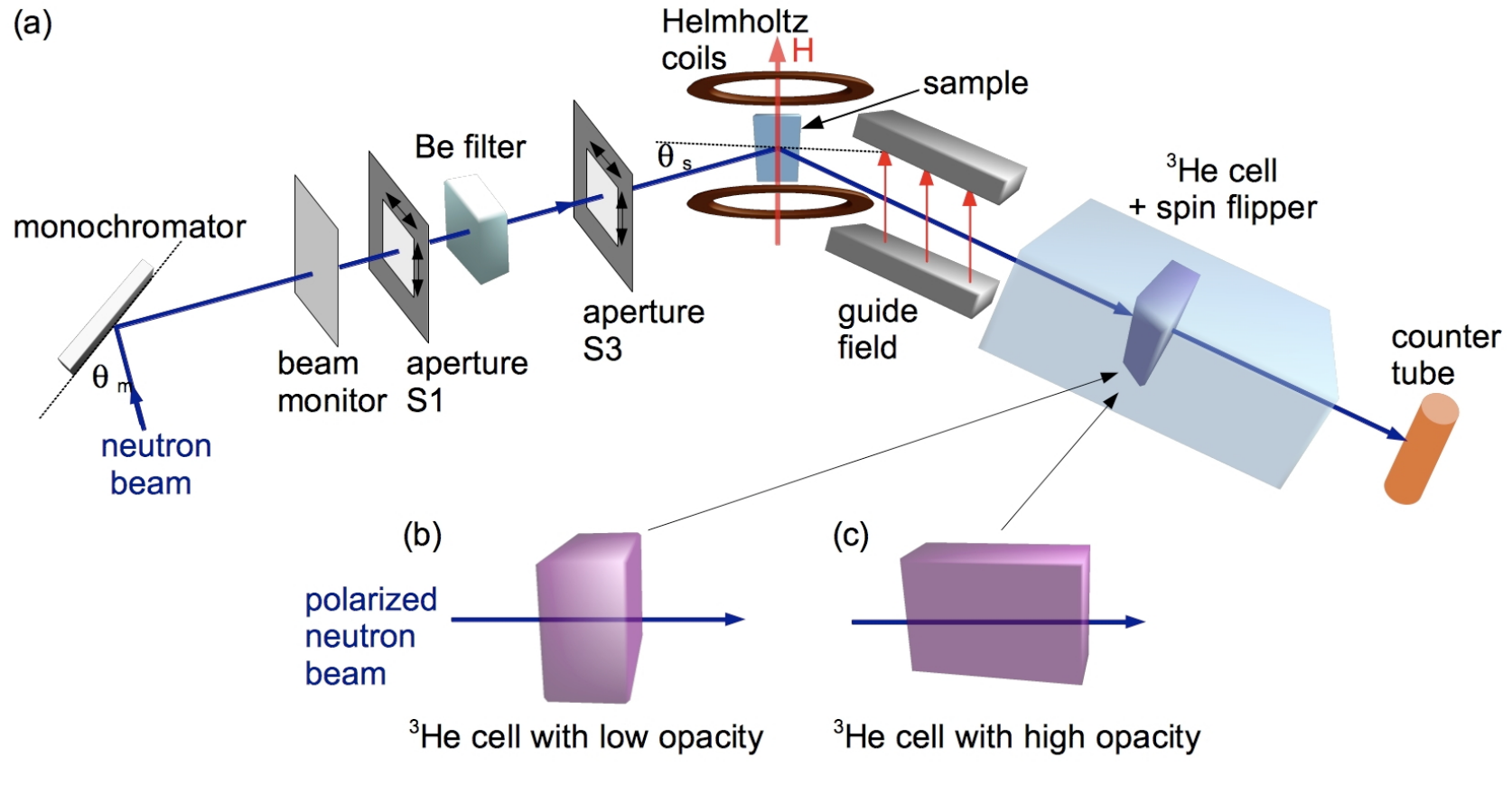}
 \caption{(a) Set-up for the polarization analysis of the Cu$_{2}$MnAl crystals at the MIRA2 beamline at FRM\,II. Details are given in the text. Arrangement of the $^3$He cell (b) with low opacity and (c) with high opacity. }
 \label{fig:Mira-Pol-Aufbau}
\end{figure}

The measurements were carried out at the MIRA2 beamline at FRM\,II. The set-up used for polarization analysis is shown in  Fig.\,\ref{fig:Mira-Pol-Aufbau}\,(a). An important prerequisite for polarization analysis is a continuous magnetic field along the flight path of the polarized neutrons since strong field gradients and especially zero field crossings lead to a depolarization of the neutron beam.

The cross section of the monochromatized neutron beam of wavelength $\lambda$\,=\,4.2$\,\pm$\,0.1\,{\AA} was determined by the source aperture (S1) and the sample aperture (S3). In this experiment the apertures were S1\,=\,3\,$\times$\,4\,mm$^2$ and S3\,=\,5\,$\times$\,8\,mm$^2$ (width$\times$height), ensuring that the small samples were entirely illuminated by the neutron beam. A Be filter at a temperature of 30\,K was used to remove neutrons with higher order wavelengths. The sample was positioned on a goniometer in an external magnetic field. A magnetic guide field provided a continuous magnetic field for the neutron beam after the (111) Bragg scattering at the sample. 

The polarization of the neutron beam was analyzed with a $^3$He cell that was provided by the HELIOS group of the FRM\,II\,\cite{hel}. The $^3$He cell was placed inside a magnetic cavity that acts as a guide field. The cavity furthermore allowed to flip the polarization of the $^3$He gas with an integrated adiabatic fast passage (AFP) flipper device\,\cite{bab07}. A $^3$He counter tube downstream of the cavity was used as a detector.

The polarization analysis was carried out with two different arrangements of the $^{3}$He cell as shown in Fig.\,\ref{fig:Mira-Pol-Aufbau}\,(b, c). For the first set of measurements the $^{3}$He cell was positioned perpendicular to the neutron beam. With this set-up the flight path of the neutrons through the polarized $^{3}$He gas is short and the absorption is reduced, leading to a low opacity. In this configuration the $^{3}$He cells had a polarization efficiency $P_{0}^{\rm rel}$\,$\sim$\,80\,-\,85\%\,\cite{mas10}. This only allows a relative measure of the polarization efficiency of the Cu$_{2}$MnAl crystals, but leads to good counting statistics and was therefore chosen to test the experimental set-up and record rocking scans and field dependencies. For an absolute measure of the polarization efficiency of the crystals, the $^{3}$He cell was positioned parallel to the neutron beam. With this set-up the flight path of the neutrons through the polarized $^{3}$He gas is long, hence leading to a high opacity. In this configuration the $^{3}$He cells had a polarization efficiency of $P_{0}^{\rm abs}$\,$>$\,99\%\,\cite{mas10}.

\begin{figure}[b]
\centering
\includegraphics[width=0.65\columnwidth]{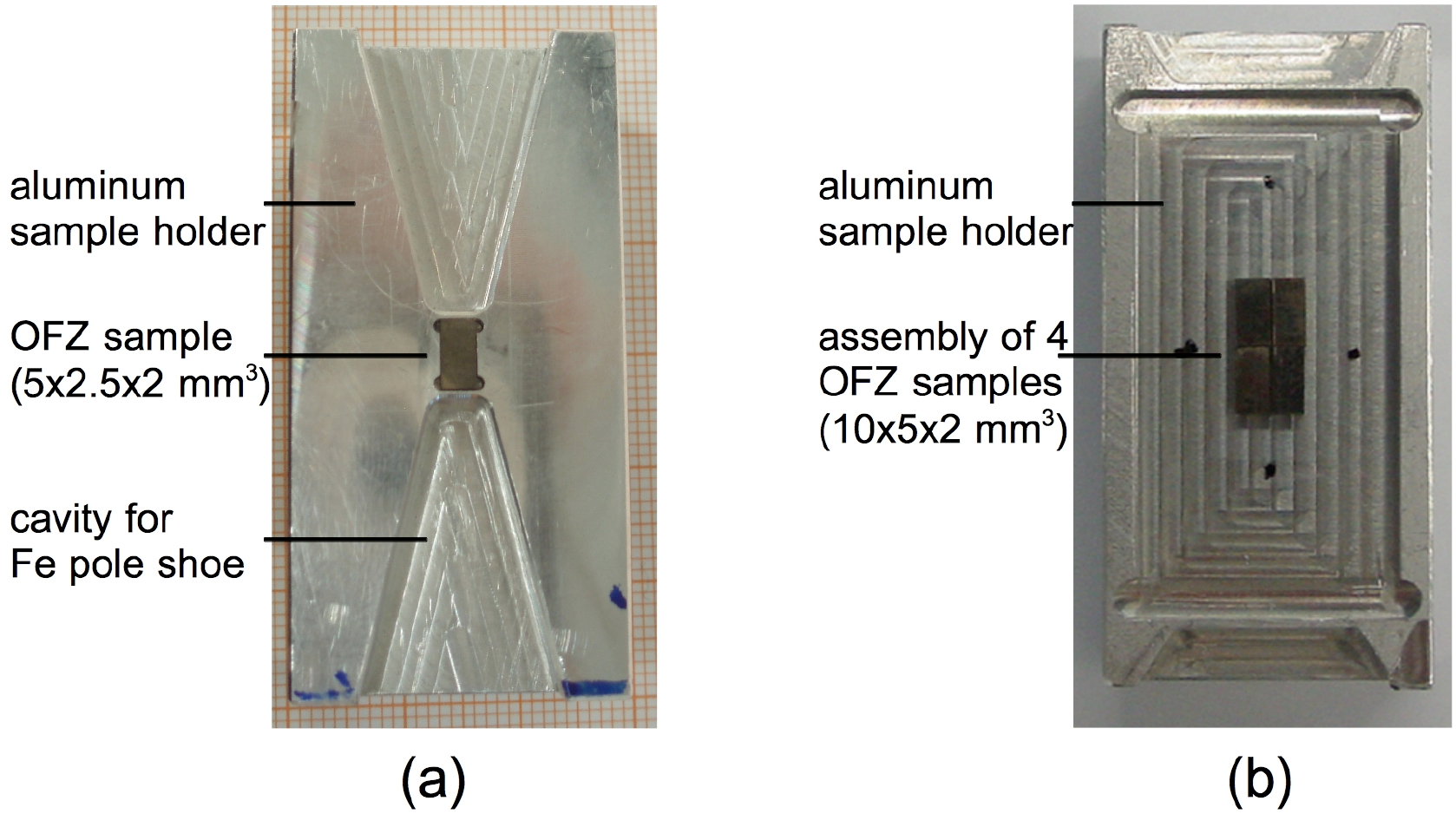}
\caption{(a) Floating zone grown Cu$_{2}$MnAl sample mounted in a bespoke aluminum holder. (b) Assembly of four OFZ crystals as arranged for the polarization analysis.}
\label{fig:ProbenMira}
\end{figure}

The samples grown in the image furnace (OFZ3, OFZ5, OFZ6 and OFZ10) and the small Bridgman grown sample (BMsmall) investigated were the same as those used for the magnetization measurements. These samples were small with dimensions of 5\,$\times$\,2.5\,$\times$\,2\,mm$^{3}$.  In addition, the large Bridgman grown (BM) crystal (40\,$\times$\,20\,$\times$\,3\,mm$^{3}$) that was characterized at HEIDI and a large inhomogeneously shaped slab (BMlarge, in average $60\times30\times$4\,mm$^{3}$), from which BMsmall was cut, were investigated. All samples were prepared and mounted with the large front side being a (111) plane. 

In a first test the OFZ samples were mounted in a bespoke aluminum holder as shown in Fig.\,\ref{fig:ProbenMira}\,(a). The holder was clamped within a horseshoe magnet where additional Fe pieces served as pole shoes. This set-up (without the Helmholtz coils) provided a magnetic field of $\sim$\,180\,mT, which is twice the field necessary to saturate the samples (cf.  Fig.\,\ref{fig:Cu2MnAl-MVergleich}). 

However, measurements with this set-up resulted in unexpected low flipping ratios of $R$\,$\sim$\,2. We believe that field gradients surrounding the sample lead to a depolarization of the beam right after the scattering process and, hence, to the low flipping ratios observed.
Therefore, the set-up was changed. The horseshoe magnet and pole shoes were removed, and instead a homogeneous magnetic field was generated by a set of Helmholtz coils as shown in Fig.\,\ref{fig:Mira-Pol-Aufbau}\,(a). With the Helmholtz coils a magnetic field of up to 220\,mT could be applied.  

Rocking scans were recorded at a 2$\theta$ angle of 74.2$^{\circ}$, appropriate for the (111) Bragg reflex of a cubic crystal structure with lattice constant $a$\,=\,5.996\,{\AA} ($d_{(111)}$\,=\,$a/\sqrt3$) and a neutron wavelength $\lambda$\,=\,4.2\,{\AA}. Typically the sample was rocked through a range of 3$^{\circ}$.

\subsubsection{$^{3}$He cell with low opacity}

\begin{figure}[p]
\centering
\includegraphics[width=0.45\columnwidth]{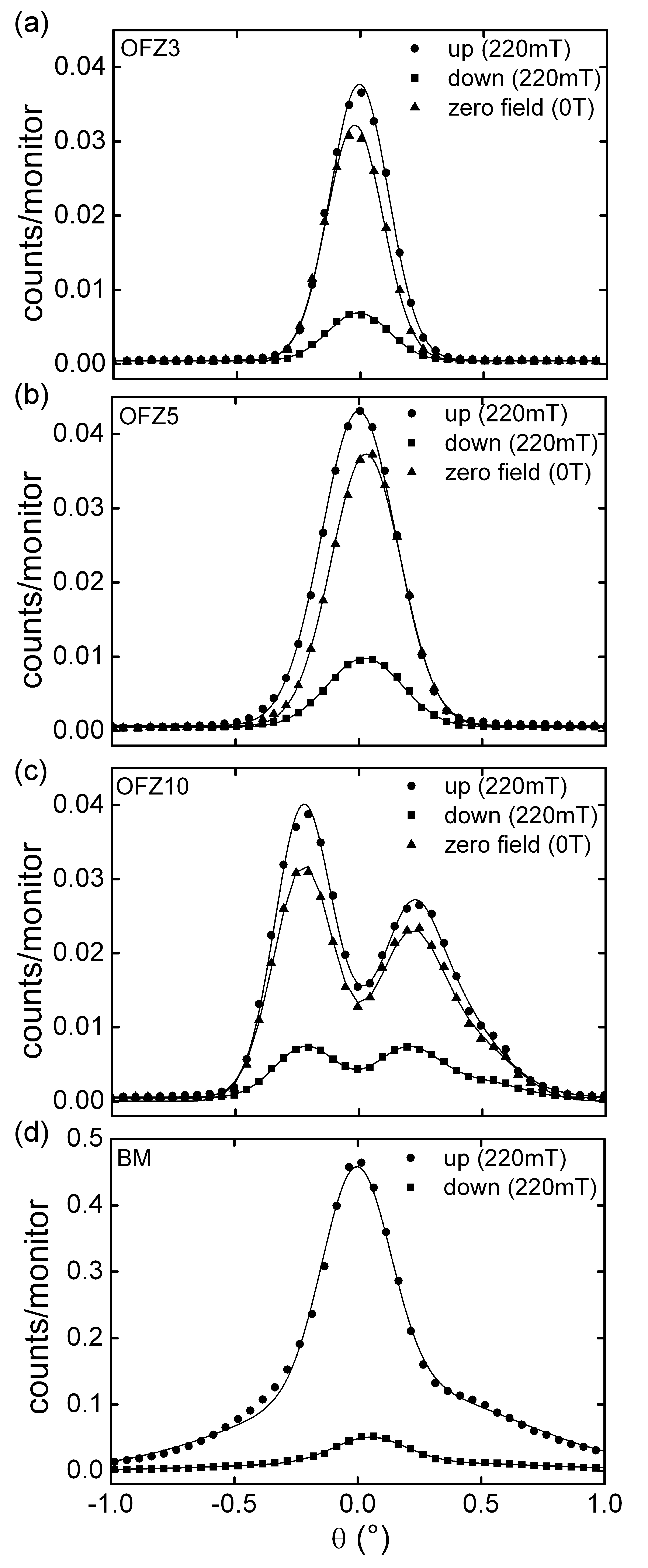}
 \caption{Rocking scans at the (111) Bragg reflexion of floating zone grown crystals OFZ3, OFZ5 and OFZ10 and of the Bridgman grown crystal (BM) with the $^{3}$He cell perpendicular to the neutron beam. Measurements were taken in spin-up ($I_{+}$) and spin-down ($I_{-}$) configuration in an external field of 220\,mT and in zero field for the OFZ crystals. The intensities are normalized to counts per monitor values. Note that the intensities for the large BM crystal are around 10 times larger than those for the small OFZ crystals. }
 \label{fig:PolOmegaScans}
\end{figure}

Rocking scans in an applied field of 220\,mT were recorded analyzing both spin configurations: the spin-up configuration ($I_{+}$, Eq.\,\ref{Iplus}), for which the $^{3}$He cell allows neutrons to pass with spin parallel to magnetization direction; and the spin-down configuration ($I_{-}$, Eq.\,\ref{Iminus}), for which the $^{3}$He cell allows neutrons to pass with spin antiparallel to the magnetization direction. In addition, rocking scans with no applied field were recorded. The results for crystals OFZ3, OFZ5, OFZ10 and BM are shown in  Fig.\,\ref{fig:PolOmegaScans}. The curves are Gaussian fits to the data.

Similar to the results of the measurements at HEIDI, a narrow (111) Bragg peak was observed for OFZ3 and OFZ5, as well as the double peak structure for OFZ10 and the slightly broadened peak for the Bridgman grown sample (BM). As expected from Eq.\,\ref{Izf}$-$\ref{Iminus} the maximum intensity was obtained for the spin-up configuration, the minimum intensity for the spin-down configuration and an intensity maximum close to the spin-up configuration for the zero field measurements. Analysis of the maximum intensities gives a flipping ratio $R$\,$\sim$\,4.5 for OFZ3, OFZ5 and OFZ10 and a flipping ratio of $R$\,$\sim$\,10 for the Bridgman crystal.

\begin{figure}[t]
\centering
\includegraphics[width=0.5\columnwidth]{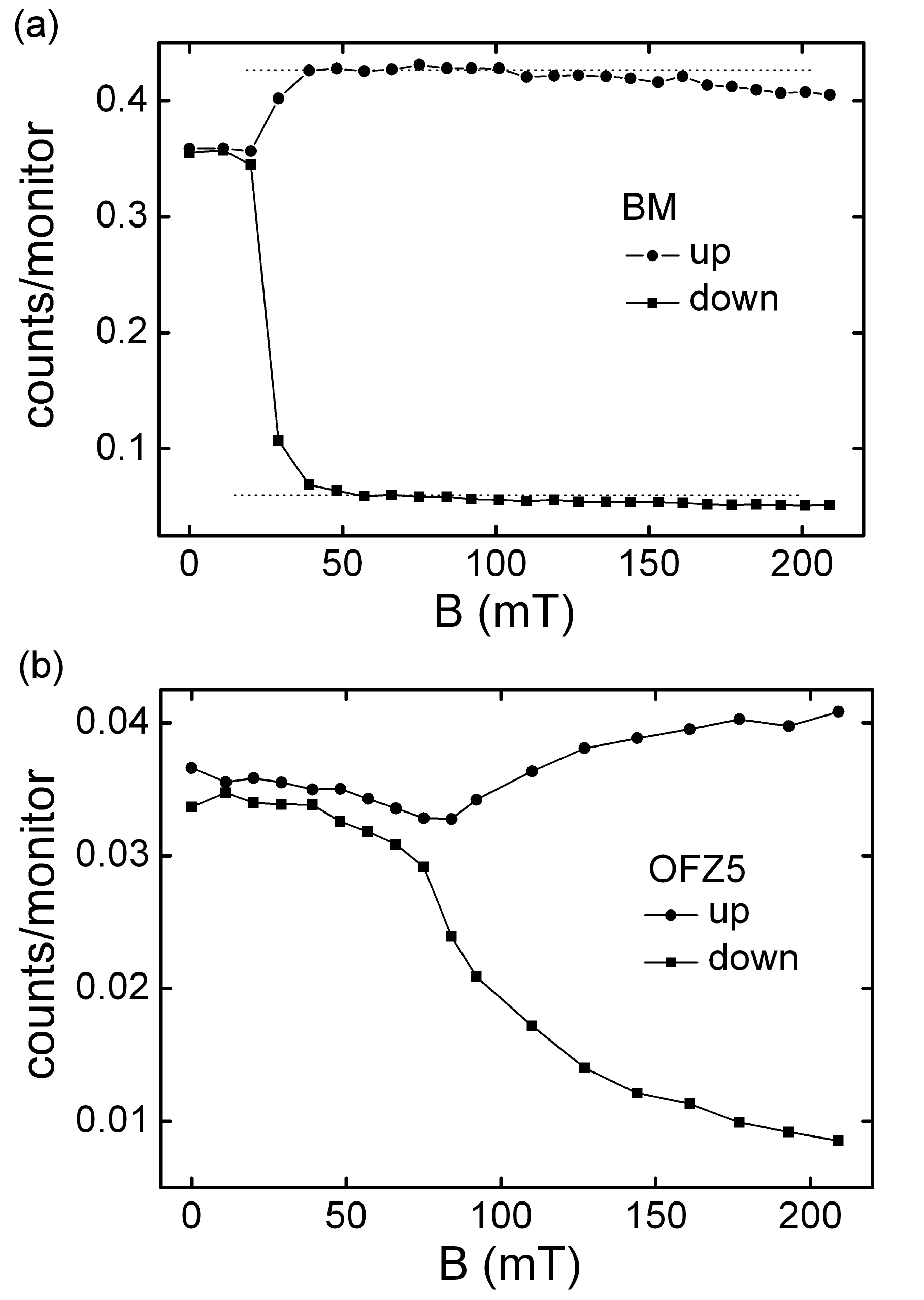}
 \caption{Field dependence of the Bragg peak maximum for the spin-up and the spin-down configuration. (a) For the large Bridgman grown crystal the intensities for both spin configurations start at a similar value, split step-like at a field above 20\,mT and saturate at fields exceeding 40\,mT. (b) Different behavior is observed for the floating zone grown crystal OFZ5. With increasing field both intensities slightly decrease and split for fields above 75\,mT. The splitting is gradual with no clear saturation up to a field of 220\,mT.}
 \label{fig:PolFieldScans}
\end{figure}

Further, the field dependence of the maximum Bragg intensity of crystals OFZ5 and BM, both for the spin-up and the spin-down configuration, was investigated. As shown in Fig.\,\ref{fig:PolFieldScans}, the two crystals show different behavior.  For the large BM crystal both intensities remain at the same level for fields below 20\,mT. With increasing field the spin-up intensity rises towards its maximum value at around 40\,mT and saturates. The spin-down intensity shows a strong decrease above 20\,mT and saturates at low intensities for fields above 40\,mT. This behavior is in agreement with Eq.\,\ref{Iplus} and Eq.\,\ref{Iminus}, if  a ferromagnetic saturation of the sample around 40\,mT is assumed. This is plausible due to its elongated form and, hence, reduced demagnetization factor compared to the OFZ crystals. The slight decrease of both intensities at the highest fields may be caused by inhomogeneous field distributions around the crystals.  

Different behavior is observed for the floating zone grown crystal OFZ5 (see Fig.\,\ref{fig:PolFieldScans}\,(b)). Here spin-up and spin-down intensities start at similar intensities at zero field. With increasing fields up to 75\,mT both intensities slightly decrease. At 75\,mT the curves split and show a curved increase (spin-up) and decrease (spin-down) to higher fields. No clear saturation is observed for fields up to 220\,mT. This behavior is in stark contrast to what was expected from the magnetization curves (cf. Fig.\,\ref{fig:Cu2MnAl-MVergleich}).  At a field of 90\,mT both the spin-up and spin-down intensities were expected to saturate. We believe that the small sample dimension is responsible for the unconventional field dependence of the intensities and, hence, for the low flipping ratio. The geometry of the sample might reduce the magnetic polarization of the sample and generate inhomogeneous field distributions inside and outside the sample that depolarize the scattered neutrons. 

\begin{table*}
\begin{center}
\begin{tabular}{|>{\centering\arraybackslash}m{3cm}|>{\centering\arraybackslash}m{1.7cm}|>{\centering\arraybackslash}m{1.7cm}|>{\centering\arraybackslash}m{1.7cm}|>{\centering\arraybackslash}m{1.7cm}|>{\centering\arraybackslash}m{1.7cm}|}
\hline
\textsc{Crystal}&\textsc{OFZ3}&\textsc{OFZ5}&\textsc{OFZ6}&\textsc{OFZ10}&\textsc{BM}\\
\hline\hline
Sample dimension \newline(mm$^{3}$)& 5\,$\times$\,2.5\,$\times$\,2    & $5\times2.5\times2 $   &$5\times2.5\times2$     &$5\times2.5\times2 $   &$40\times20\times3 $ \\
\hline
Flipping ratio $R$& 8.6  & 8.3 & 8.1 & 9.8/5.2 & 60.3 \\
\hline
Polarization efficiency $P$ (\%)& 79 & 79 & 78 & 81/68 & 97 \\
\hline
\end{tabular}
\end{center}
\caption{Sample dimensions, flipping ratios and polarization efficiencies for the small floating zone crystals (OFZ) and the large Brigdman grown crystal (BM). }
\label{tab:mira2}
\end{table*}

\subsubsection{$^{3}$He cell with high opacity}
In order to measure the absolute polarization, the last set of data was taken with the $^{3}$He cell parallel to the neutron beam (see  Fig.\,\ref{fig:Mira-Pol-Aufbau}\,(c)).  The intensity at the maximum Bragg peak positions of each sample for both spin-up and spin-down configuration was recorded as well as the background. The measurement times were increased in order to obtain good counting statistics. 
The resulting flipping ratios and polarization efficiencies are shown in Table\,\ref{tab:mira2}. A low polarization efficiency of $P$\,$\sim$\,80\% for the small floating zone grown crystals was obtained as compared to the very good $P$\,$\sim$\,97\% for the large Bridgman grown crystal. 

\subsubsection{Role of sample geometry}

In order to investigate the influence of the sample geometry on the polarization efficiency two independent measurements were performed. First, a large Bridgman grown sample (BMlarge) and a small sample (BMsmall) prepared from BMlarge with dimensions similar to the OFZ crystals were examined. As a second test an assembly of the floating zone crystals OFZ3, OFZ5, OFZ6 and OFZ10, as shown in  Fig.\,\ref{fig:ProbenMira}\,(b), was measured.

From the first measurements we obtained a flipping ratio $R$\,=\,33 for  BMlarge in comparison to $R$\,=\,8.2 for BMsmall. The flipping ratio $R$\,=\,8.2 for the small Bridgman grown sample is similar to the low value obtained for the small floating zone grown crystals. 

\begin{figure}[h]
\centering
\includegraphics[width=0.45\columnwidth]{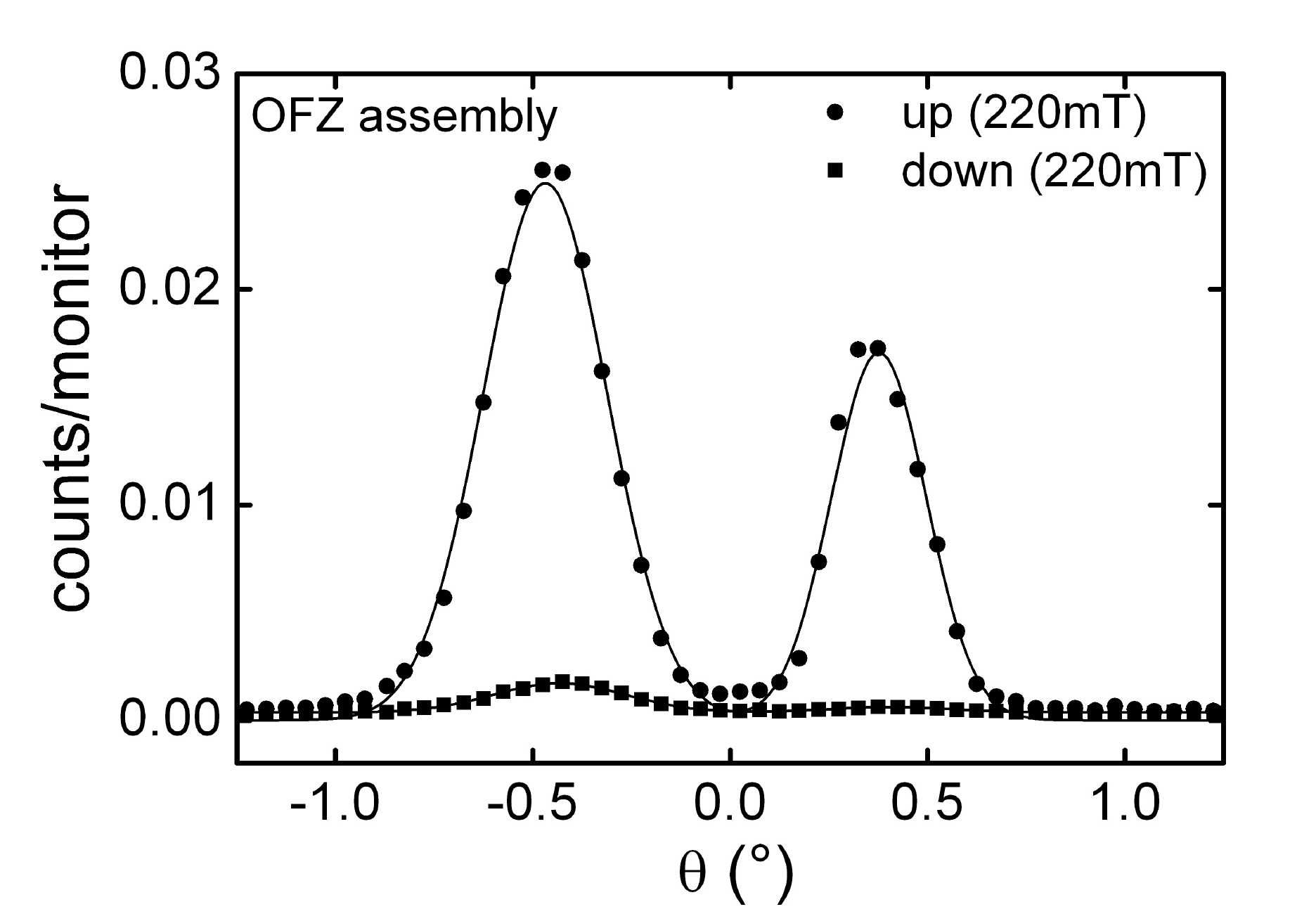}
 \caption{Rocking scan of the assembly of floating zone crystals OFZ3, OFZ5, OFZ6 and OFZ10 for both spin-up and spin-down configuration. Misalignment of the (111) planes of each crystal with respect to each other leads to the double peak structure. The assembly shows increased flipping ratios of $R=23$ and $R=20$ for each of the two Bragg peaks compared to $R$\,$\sim$\,8 for each OFZ sample by itself.}
 \label{fig:Mosaic-Dunkelzelle}
\end{figure}

The rocking scan of the assembly, as the second test, is shown in  Fig.\,\ref{fig:Mosaic-Dunkelzelle}. The rocking curve shows a clear double peak structure that results from a slight misalignment of the (111) planes of each crystal. Nevertheless, the flipping ratios of each of the two Bragg peaks are 23 and 20 (see Table\,\ref{tab:mira3}).

The increase of the flipping ratios from $R$\,$\sim$\,8 of each OFZ sample by itself to $R$\,$\sim$\,20 for an assembly of the same samples and the reduced flipping ratio $R$\,=\,8.2 of the small Bridgman sample compared to $R$\,=\,33 of the large Bridgman sample clearly identify the sample dimension as the origin of the low flipping ratios and polarization efficiencies of the floating zone grown crystals. Since the small floating zone crystals and the small Bridgman sample show a similar flipping ratio $R$\,$\sim$\,8,  the floating zone grown crystals are expected to achieve high polarization efficiencies comparable to those obtained for Bridgman grown crystals if the problems that arise due to the sample geometry are avoided. This may be achieved  by growing larger crystals, by an assembly of several samples, or by a suppression of the stray fields or field inhomogeneities by embedding the Cu$_{2}$MnAl crystals in an adequate ferromagnetic material.

\begin{table}[h!]
\begin{center}
\begin{tabular}{|>{\centering\arraybackslash}m{3,5cm}|>{\centering\arraybackslash}m{2cm}|>{\centering\arraybackslash}m{2.2cm}|>{\centering\arraybackslash}m{2.1cm}|}
\hline
\textsc{Crystal}&\textsc{OFZ assembly}&\textsc{BM large}&\textsc{BM small}\\
\hline\hline
Sample dimension (mm$^{3}$)& $10\times4\times2  $ &$\sim60\times30\times4$    & $5\times2.4\times1.8$   \\
\hline
Flipping ratio $R$& 23/20 & 33 & 8.2 \\
\hline
Polarization efficiency $P$(\%)& 92/90 & 94 & 79\\
\hline
\end{tabular}
\end{center}
\caption{Sample dimensions, flipping ratios and polarization efficiencies of Cu$_{2}$MnAl samples OFZassembly, BMlarge and BMsmall additionally measured at MIRA in order to  analyze the dependence of the polarization efficiency on the sample geometry.}
\label{tab:mira3}
\end{table}

\section{\label{sec:level6}Conclusion}
 
In summary eight single crystals of the Heusler compound Cu$_{2}$MnAl were grown by crucible-free floating zone: two in a vertical double ellipsoid image furnace at IFW Dresden and six with a UHV-compatible image furnace at TUM. We found that Cu$_{2}$MnAl shows a strong tendency to crystallize in the cubic $\rm{L2_{1}}$ crystal structure, indicating a congruent melting formation. The temperature gradient of the image furnace seems to be large enough to avoid the decomposition of Cu$_{2}$MnAl. The high purity static inert gas environment in the UHV-compatible image furnace at TUM was indispensable to reduce the oxygen formation around the molten zone. High purity conditions allowed to establish stable growth conditions and, hence, to obtain single crystals across the entire diameter for each of the six crystals grown with the image furnace at TUM. No preferred growth direction of the floating zone grown crystals could be identified.  

Comparison of the magnetic properties of four crystals established a reproducible magnetic moment of $m$\,$\sim$\,3.6\,$\mu_{\rm B}$/f.u. at 4\,K and $m$\,$\sim$\,3.2\,$\mu_{\rm B}$/f.u. at 300\,K for each crystal (within 3\%). This is in perfect agreement with the magnetic moments measured for a Bridgman grown crystal and those reported in literature\,\cite{mic78}. 

Neutron diffraction of the  $\{400\}$ and $\{111\}$ Bragg intensities established an isotropic mosaic spread of the floating zone grown crystals when constant growth parameters were applied. This is a clear advantage compared to Bridgman grown crystals where an anisotropic mosaic spread is reported\,\cite{cou99}. 

Further, a study of the polarizing properties was performed. For a large Bridgman grown crystal a high polarization efficiency of 97\% was found. The lower polarization efficiency of the floating zone grown crystals was found to be due to their small sample dimensions and could be raised to 91\% by an assembly of four small crystals.

In conclusion, single crystal growth of Cu$_{2}$MnAl with optical floating zone allows to reproducibly grow crystals with a homogeneous mosaic distribution. This avoids the main drawback of the Bridgman grown crystals\,\cite{cou99}. For commercial applications it will be necessary to grow single crystals with a larger diameter. Moreover, a seed with a predefined orientation may allow to prepare larger samples with a (111) plane from the crystals grown. Actually, recent growth experiments already allowed the successful growth of oriented Cu$_{2}$MnAl single crystals with a diameter of up to 10\,mm\,\cite{bau12}. Since the size of polarizing crystals typically used for technical applications in neutron scattering starts at around 10\,$\times$\,20\,mm$^{2}$\,\cite{bon10}, we believe that in future investigations these dimensions should be accessible with floating zone crystal growth.

\section*{Acknowledgments}
We gratefully acknowledge support and discussions with S. Masalovitch, B. Russ, T. Adams, S. M\"uhlbauer, N. Vizent, J. Repper, A. Mantwill, R. Schwikowski, the team of the crystal laboratory at TUM, and the team of FRM II. We also gratefully acknowledge financial support through the DFG collaborative research  network TRR80 (From electronic correlations to functionality), the DFG research unit FOR960 (quantum phase transitions) and individual grants under contract PF393/10 and PF393/11.



\begin{thebibliography}{28}
\expandafter\ifx\csname natexlab\endcsname\relax\def\natexlab#1{#1}\fi
\expandafter\ifx\csname bibnamefont\endcsname\relax
  \def\bibnamefont#1{#1}\fi
\expandafter\ifx\csname bibfnamefont\endcsname\relax
  \def\bibfnamefont#1{#1}\fi
\expandafter\ifx\csname citenamefont\endcsname\relax
  \def\citenamefont#1{#1}\fi
\expandafter\ifx\csname url\endcsname\relax
  \def\url#1{\texttt{#1}}\fi
\expandafter\ifx\csname urlprefix\endcsname\relax\def\urlprefix{URL }\fi
\providecommand{\bibinfo}[2]{#2}
\providecommand{\eprint}[2][]{\url{#2}}

\bibitem[{\citenamefont{Chadov et~al.}(2010)\citenamefont{Chadov, Qi, K\"ubler,
  Fecher, Felser, and Zhang}}]{cha10}
\bibinfo{author}{\bibfnamefont{S.}~\bibnamefont{Chadov}},
  \bibinfo{author}{\bibfnamefont{X.}~\bibnamefont{Qi}},
  \bibinfo{author}{\bibfnamefont{J.}~\bibnamefont{K\"ubler}},
  \bibinfo{author}{\bibfnamefont{G.~H.} \bibnamefont{Fecher}},
  \bibinfo{author}{\bibfnamefont{C.}~\bibnamefont{Felser}}, \bibnamefont{and}
  \bibinfo{author}{\bibfnamefont{S.-C.} \bibnamefont{Zhang}},
  \bibinfo{journal}{Nature Materials} \textbf{\bibinfo{volume}{9}},
  \bibinfo{pages}{541} (\bibinfo{year}{2010}).

\bibitem[{\citenamefont{Graf et~al.}(2011)\citenamefont{Graf, Felser, and
  Parkin}}]{graf2011}
\bibinfo{author}{\bibfnamefont{T.}~\bibnamefont{Graf}},
  \bibinfo{author}{\bibfnamefont{C.}~\bibnamefont{Felser}}, \bibnamefont{and}
  \bibinfo{author}{\bibfnamefont{S.~S.} \bibnamefont{Parkin}},
  \bibinfo{journal}{Progress in Solid State Chemistry}
  \textbf{\bibinfo{volume}{39}},
  \bibinfo{pages}{1-50}
  (\bibinfo{year}{2011}).
  

\bibitem[{\citenamefont{{Felser} and {Hillebrands}}(2009)}]{fel09}
\bibinfo{author}{\bibfnamefont{C.}~\bibnamefont{{Felser}}} \bibnamefont{and}
  \bibinfo{author}{\bibfnamefont{B.}~\bibnamefont{{Hillebrands}}},
  \bibinfo{journal}{Journal of Physics D Applied Physics}
  \textbf{\bibinfo{volume}{42}}, \bibinfo{pages}{080301}
  (\bibinfo{year}{2009}).

\bibitem[{\citenamefont{Sprungmann et~al.}(2010)\citenamefont{Sprungmann,
  Westerholt, Zabel, Weides, and Kohlstedt}}]{spr10}
\bibinfo{author}{\bibfnamefont{D.}~\bibnamefont{Sprungmann}},
  \bibinfo{author}{\bibfnamefont{K.}~\bibnamefont{Westerholt}},
  \bibinfo{author}{\bibfnamefont{H.}~\bibnamefont{Zabel}},
  \bibinfo{author}{\bibfnamefont{M.}~\bibnamefont{Weides}}, \bibnamefont{and}
  \bibinfo{author}{\bibfnamefont{H.}~\bibnamefont{Kohlstedt}},
  \bibinfo{journal}{Physical Review B} \textbf{\bibinfo{volume}{82}},
  \bibinfo{pages}{060505} (\bibinfo{year}{2010}).

\bibitem[{\citenamefont{Erb et~al.}(2010)\citenamefont{Erb, Nowak, Westerholt,
  and Zabel}}]{erb10}
\bibinfo{author}{\bibfnamefont{D.}~\bibnamefont{Erb}},
  \bibinfo{author}{\bibfnamefont{G.}~\bibnamefont{Nowak}},
  \bibinfo{author}{\bibfnamefont{K.}~\bibnamefont{Westerholt}},
  \bibnamefont{and} \bibinfo{author}{\bibfnamefont{H.}~\bibnamefont{Zabel}},
  \bibinfo{journal}{Journal of Physics D: Applied Physics}
  \textbf{\bibinfo{volume}{43}}, \bibinfo{pages}{285001}
  (\bibinfo{year}{2010}).

\bibitem[{\citenamefont{Heusler}(1903)}]{heu03}
\bibinfo{author}{\bibfnamefont{F.}~\bibnamefont{Heusler}},
  \bibinfo{journal}{{Verhandlungen der Deutschen Physikalischen Gesellschaft }}
  \textbf{\bibinfo{volume}{5}} (\bibinfo{year}{1903}).

\bibitem[{\citenamefont{{Michelutti}}(1978)}]{mic78}
\bibinfo{author}{\bibfnamefont{B.}~\bibnamefont{{Michelutti}}},
  \bibinfo{journal}{Solid State Communications} \textbf{\bibinfo{volume}{25}},
  \bibinfo{pages}{163} (\bibinfo{year}{1978}).

\bibitem[{\citenamefont{Williams}(1988)}]{wil88}
\bibinfo{author}{\bibfnamefont{W.~G.} \bibnamefont{Williams}},
  \emph{\bibinfo{title}{{Polarized neutrons}}} (\bibinfo{publisher}{Clarendon
  Press, Oxford, New York}, \bibinfo{year}{1988}).

\bibitem[{\citenamefont{Delapalme et~al.}(1971)\citenamefont{Delapalme,
  Schweizer, Couderchon, and de~la Bathie}}]{del71}
\bibinfo{author}{\bibfnamefont{A.}~\bibnamefont{Delapalme}},
  \bibinfo{author}{\bibfnamefont{J.}~\bibnamefont{Schweizer}},
  \bibinfo{author}{\bibfnamefont{G.}~\bibnamefont{Couderchon}},
  \bibnamefont{and} \bibinfo{author}{\bibfnamefont{R.~P.} \bibnamefont{de~la
  Bathie}}, \bibinfo{journal}{Nuclear Instruments and Methods}
  \textbf{\bibinfo{volume}{95}}, \bibinfo{pages}{589} (\bibinfo{year}{1971}).

\bibitem[{\citenamefont{Freund et~al.}(1983)\citenamefont{Freund, Pynn,
  Stirling, and Zeyen}}]{fre83}
\bibinfo{author}{\bibfnamefont{A.}~\bibnamefont{Freund}},
  \bibinfo{author}{\bibfnamefont{R.}~\bibnamefont{Pynn}},
  \bibinfo{author}{\bibfnamefont{W.}~\bibnamefont{Stirling}}, \bibnamefont{and}
  \bibinfo{author}{\bibfnamefont{C.}~\bibnamefont{Zeyen}},
  \bibinfo{journal}{Physica B\&C} \textbf{\bibinfo{volume}{120}},
  \bibinfo{pages}{86} (\bibinfo{year}{1983}).
  
  \bibitem{fre09}
\bibinfo{author}{\bibfnamefont{A. K. }~\bibnamefont{Freund}},
  \bibinfo{journal}{Journal of Applied Crystallography}
  \textbf{\bibinfo{volume}{42}},
  \bibinfo{pages}{36-42} (\bibinfo{year}{2009}).

\bibitem[{\citenamefont{Courtois}(1999)}]{cou99}
\bibinfo{author}{\bibfnamefont{P.}~\bibnamefont{Courtois}},
  \bibinfo{journal}{Physica B: Condensed Matter}
  \textbf{\bibinfo{volume}{267-268}}, \bibinfo{pages}{363}
  (\bibinfo{year}{1999}).

\bibitem[{\citenamefont{Courtois et~al.}(2004)\citenamefont{Courtois, Hamelin,
  and Andersen}}]{cou04}
\bibinfo{author}{\bibfnamefont{P.}~\bibnamefont{Courtois}},
  \bibinfo{author}{\bibfnamefont{B.}~\bibnamefont{Hamelin}}, \bibnamefont{and}
  \bibinfo{author}{\bibfnamefont{K.~H.} \bibnamefont{Andersen}},
  \bibinfo{journal}{Nuclear Instruments and Methods in Physics Research Section
  A: Accelerators, Spectrometers, Detectors and Associated Equipment}
  \textbf{\bibinfo{volume}{529}}, \bibinfo{pages}{157} (\bibinfo{year}{2004}).

\bibitem[{\citenamefont{Neubauer et~al.}(2011)\citenamefont{Neubauer, Boeuf,
  Bauer, Russ, v.~L\"{o}hneysen, and Pfleiderer}}]{neu11}
\bibinfo{author}{\bibfnamefont{A.}~\bibnamefont{Neubauer}},
  \bibinfo{author}{\bibfnamefont{J.}~\bibnamefont{Boeuf}},
  \bibinfo{author}{\bibfnamefont{A.}~\bibnamefont{Bauer}},
  \bibinfo{author}{\bibfnamefont{B.}~\bibnamefont{Russ}},
  \bibinfo{author}{\bibfnamefont{H.}~\bibnamefont{v.~L\"{o}hneysen}},
  \bibnamefont{and}
  \bibinfo{author}{\bibfnamefont{C.}~\bibnamefont{Pfleiderer}},
  \bibinfo{journal}{Review of Scientific Instruments}
  \textbf{\bibinfo{volume}{82}}, \bibinfo{pages}{013902}
  (\bibinfo{year}{2011}).

\bibitem[{\citenamefont{K\"oster and G\"odecke}(1966)}]{kos66}
\bibinfo{author}{\bibfnamefont{W.}~\bibnamefont{K\"oster}} \bibnamefont{and}
  \bibinfo{author}{\bibfnamefont{T.}~\bibnamefont{G\"odecke}},
  \bibinfo{journal}{Zeitschrift f\"ur Metallkunde}
  \textbf{\bibinfo{volume}{57}}, \bibinfo{pages}{889} (\bibinfo{year}{1966}).

\bibitem[{\citenamefont{Dubois and Cheverau}(1979)}]{dub79}
\bibinfo{author}{\bibfnamefont{B.}~\bibnamefont{Dubois}} \bibnamefont{and}
  \bibinfo{author}{\bibfnamefont{D.}~\bibnamefont{Cheverau}},
  \bibinfo{journal}{Journal of Materials Science}
  \textbf{\bibinfo{volume}{14}}, \bibinfo{pages}{2296} (\bibinfo{year}{1979}).

\bibitem[{\citenamefont{Sakka and Nakamura}(1990)}]{sak90}
\bibinfo{author}{\bibfnamefont{Y.}~\bibnamefont{Sakka}} \bibnamefont{and}
  \bibinfo{author}{\bibfnamefont{M.}~\bibnamefont{Nakamura}},
  \bibinfo{journal}{Journal of Materials Science}
  \textbf{\bibinfo{volume}{25}}, \bibinfo{pages}{2549} (\bibinfo{year}{1990}).

\bibitem[{\citenamefont{Kainuma et~al.}(1998)\citenamefont{Kainuma, Satoh, Liu,
  Ohnuma, and Ishida}}]{kai98}
\bibinfo{author}{\bibfnamefont{R.}~\bibnamefont{Kainuma}},
  \bibinfo{author}{\bibfnamefont{N.}~\bibnamefont{Satoh}},
  \bibinfo{author}{\bibfnamefont{X.~J.} \bibnamefont{Liu}},
  \bibinfo{author}{\bibfnamefont{I.}~\bibnamefont{Ohnuma}}, \bibnamefont{and}
  \bibinfo{author}{\bibfnamefont{K.}~\bibnamefont{Ishida}},
  \bibinfo{journal}{Journal of Alloys and Compounds}
  \textbf{\bibinfo{volume}{266}}, \bibinfo{pages}{191} (\bibinfo{year}{1998}).

  \bibitem[{neu()}]{neu12}
\bibinfo{note}{Seed and feed rods for crystals HKZ363, OFZ1, OFZ3, OFZ4 and OFZ5 were all prepared in the rod-casting furnace at the IFW Dresden from the same batch of Bridgman grown Cu$_2$MnAlsingle crystals. The seed and feed rods for crystals OFZ6 and OFZ10 were prepared from pure starting material in the rod-casting furnace at TUM}.

\bibitem[{\citenamefont{Neubauer}(2011)}]{neuphd11}
\bibinfo{author}{\bibfnamefont{A.}~\bibnamefont{Neubauer}}, Ph.D. thesis,
  \bibinfo{school}{{Technische Universit\"at M\"unchen}}
  (\bibinfo{year}{2011}).

\bibitem[{\citenamefont{M\"unzer}(2009)}]{mun09}
\bibinfo{author}{\bibfnamefont{W.}~\bibnamefont{M\"unzer}},
  \bibinfo{type}{Diploma thesis}, \bibinfo{school}{{Technische Universit\"at
  M\"unchen}} (\bibinfo{year}{2009}).

\bibitem[{\citenamefont{Bauer}(2009)}]{bau09}
\bibinfo{author}{\bibfnamefont{A.}~\bibnamefont{Bauer}}, \bibinfo{type}{Diploma
  thesis}, \bibinfo{school}{{Technische Universit\"at M\"unchen}}
  (\bibinfo{year}{2009}).

\bibitem[{\citenamefont{Meven et~al.}(2007)\citenamefont{Meven, Hutanu, and
  Heger}}]{heidi}
\bibinfo{author}{\bibfnamefont{M.}~\bibnamefont{Meven}},
  \bibinfo{author}{\bibfnamefont{V.}~\bibnamefont{Hutanu}}, \bibnamefont{and}
  \bibinfo{author}{\bibfnamefont{G.}~\bibnamefont{Heger}},
  \bibinfo{journal}{Neutron News} \textbf{\bibinfo{volume}{18}},
  \bibinfo{pages}{19} (\bibinfo{year}{2007}).

\bibitem[{res()}]{res}
\bibinfo{note}{The FWHM values were obtained by means of the measurement
  software at HEIDI through linear interpolation of the measured intensities. The steep edge and the small statistical error of the supporting points in the peak area give $\pm0.05^{\circ}$ as a pessimistic assumption of the instrumental error (due to the step width of 0.1$^{\circ}$). The instrumental resolution function depends on the diffraction angle 2$\theta$,
  giving a beam divergence of around 0.13$^{\circ}$ for $\{400\}$, around 0.11 for
  $\{333\}$, and around 0.34$^{\circ}$ for the $\{111\}$ scattering planes. The
  crystal mosaicities were then obtained via the relation: $
  \rm(mosaicity)^{2}=(FWHM)^{2}-(beam\,\, divergence)^{2}$}.

\bibitem[{\citenamefont{Bacon and Lonsdale}(1953)}]{bac53}
\bibinfo{author}{\bibfnamefont{G.~E.} \bibnamefont{Bacon}} \bibnamefont{and}
  \bibinfo{author}{\bibfnamefont{K.}~\bibnamefont{Lonsdale}},
  \bibinfo{journal}{{Reports on Progress in Physics}}
  \textbf{\bibinfo{volume}{16}}, \bibinfo{pages}{1} (\bibinfo{year}{1953}).

\bibitem[{\citenamefont{Webster}(1969)}]{web69}
\bibinfo{author}{\bibfnamefont{P.~J.} \bibnamefont{Webster}},
  \bibinfo{journal}{{Contemporary Physics}} \textbf{\bibinfo{volume}{10}},
  \bibinfo{pages}{559} (\bibinfo{year}{1969}).

\bibitem[{hel()}]{hel}
\bibinfo{note}{{http://www.frm2.tum.de/en/science/service-groups/neutrons-opti%
cs/3he-polarizer-helios, 2010}}.

\bibitem[{\citenamefont{Babcock et~al.}(2007)\citenamefont{Babcock, Petoukhov,
  Chastagnier, Jullien, Leli\'evre-Berna, Andersen, Georgii, Masalovich, Boag,
  Frost et~al.}}]{bab07}
\bibinfo{author}{\bibfnamefont{E.}~\bibnamefont{Babcock}},
  \bibinfo{author}{\bibfnamefont{A.}~\bibnamefont{Petoukhov}},
  \bibinfo{author}{\bibfnamefont{J.}~\bibnamefont{Chastagnier}},
  \bibinfo{author}{\bibfnamefont{D.}~\bibnamefont{Jullien}},
  \bibinfo{author}{\bibfnamefont{E.}~\bibnamefont{Leli\'evre-Berna}},
  \bibinfo{author}{\bibfnamefont{K.}~\bibnamefont{Andersen}},
  \bibinfo{author}{\bibfnamefont{R.}~\bibnamefont{Georgii}},
  \bibinfo{author}{\bibfnamefont{S.}~\bibnamefont{Masalovich}},
  \bibinfo{author}{\bibfnamefont{S.}~\bibnamefont{Boag}},
  \bibinfo{author}{\bibfnamefont{C.}~\bibnamefont{Frost}},
  \bibnamefont{et~al.}, \bibinfo{journal}{{Physica B: Condensed Matter}}
  \textbf{\bibinfo{volume}{397}}, \bibinfo{pages}{172} (\bibinfo{year}{2007}).

\bibitem[{mas()}]{mas10}
\bibinfo{note}{{Sergey Masalovitch, personal communication, 2010}}.

\bibitem[{bau()}]{bau12}
\bibinfo{note}{A. Bauer, C. Pfleiderer, unpublished}.

\bibitem[{bon()}]{bon10}
\bibinfo{note}{Peter B\"oni, personal communication, 2010}.


\end{thebibliography}
\end{document}